\documentclass[runningheads,envcountsame]{llncs}

\usepackage[T1]{fontenc}
\usepackage{lmodern}
\usepackage{amsmath,amssymb}
\usepackage{graphicx}
\usepackage{booktabs}
\usepackage{xcolor}
\usepackage{tikz}
\usetikzlibrary{arrows.meta,bending}
\usepackage{subcaption}
\usepackage{hyperref}
\hypersetup{pdfborder={0 0 0}}

\newcommand{\IE}{\textsc{Infinite Energy}}
\newcommand{\UIE}{\textsc{Unconstrained-Infinite Energy}}
\newcommand{\HP}{\textsc{Hamiltonian Path}}
\newcommand{\MAJ}{\textsc{Majority}}
\newcommand{\cc}[1]{\textsf{#1}}
\newcommand{\NP}{\cc{NP}}
\newcommand{\PSPACE}{\cc{PSPACE}}

\newcommand{\cP}{\cc{P}}
\newcommand{\cL}{\cc{L}}
\newcommand{\TC}{\cc{TC}$^0$}
\newcommand{\AC}{\cc{AC}$^0$}
\newcommand{\mTC}{\mathsf{TC}^0}
\newcommand{\mAC}{\mathsf{AC}^0}
\newcommand{\R}{\mathbb{R}}
\newcommand{\en}{\mathrm{en}}
\newcommand{\sus}{\mathrm{sus}}
\newcommand{\col}{\mathrm{col}}
\newcommand{\Col}{\mathsf{Col}}
\newcommand{\wdash}{\mkern1mu{-}\mkern1mu}

\spnewtheorem*{ieproblem}{Problem}{\bfseries}{\itshape}

\tikzset{
  board/.style={draw, line width=0.7pt, fill=black!4, rounded corners=2pt},
  sock/.style={draw, line width=0.6pt, fill=black!12},
  wsock/.style={draw, line width=0.6pt, fill=white},
  src/.style={draw, line width=0.7pt, fill=black!8},
  wire/.style={line width=1pt, arrows={-Stealth[length=4pt,width=3pt]}},
  impl/.style={line width=0.7pt, black!55, dash pattern=on 2.99pt off 1.99pt,
               arrows={-Stealth[length=4pt,width=3pt]}},
}

\begin{document}

\title{Hardness of ``Infinite Energy'' Puzzles}
\author{Utkarsh Tiwari\inst{1, 2} \and
Utkarsh Thanvi\inst{2} \and
Armaan Sethi\inst{2} \and
Aalok Thakkar\inst{3,5} \and
Neeldhara Misra\inst{4}}
\authorrunning{U. Tiwari et al.}
\institute{%
$^{1}$TTIC \quad
$^{2}$BITS Pilani \quad
$^{3}$Ashoka University \quad
$^{4}$IIT Gandhinagar \quad
$^{5}$Alt Earth}

\maketitle

\begin{abstract}
\IE{} (by Light Potato, 2025) is a single-player puzzle in which one wires a
network of sockets so that, after every power source is unplugged, energy
keeps circulating and leaves every board energized. We study several variants
of the game and the complexity of deciding whether an instance can be won. The
variants can be played at
\url{https://infinite-energy-peer-review.netlify.app}, which also has worked
examples of the reduction techniques.

The colorless game, with connectability given by an arbitrary matrix, is
\NP-hard (by reduction from \HP) and lies in \PSPACE. We give a seven-socket
instance winnable only by charging a temporary battery loop and then breaking
it, so the natural ``charge and keep'' strategy fails. We reduce
\NP-membership to one conjecture: that every winnable instance has a winning
play in which no socket is charged twice.

Two variants bound the complexity more tightly. The unconstrained variant, in
which connectability is unrestricted, is \TC-complete under \AC{} Turing
reductions (hence in \cL). The geometric variant, in which connectability is
set by proximity of points in $\R^k$, stays \NP-hard for every fixed $k \ge 2$
via reduction from \HP{} over grid graphs. The one-dimensional case is open.
The colored variant is \NP-hard, though a matching upper bound remains open.

\end{abstract}

\section{Introduction}\label{sec:intro}

\IE{} (by Light Potato, 2025) is a puzzle in which the player must assemble a
circuit that stays powered after its power supply has been taken away. It was
created for the 2025 GMTK Game Jam, an annual competition hosted by the game
designer Mark Brown. Out of the 9,547 entries it ranked first as Brown's
personal favorite and tenth in the community vote for creativity. The original
game can be played at \url{https://lightpotato.itch.io/infinite-energy}. We
study it not only for this reception but also because its complexity sits in an unusual position among well-studied games, i.e., while a lower bound on hardness follows from reductions from known \NP-hard problem, an upper bound on hardness remains elusive. To our knowledge, \textsc{Infinite Energy} does not fit cleanly into any
family of puzzles whose complexity has been studied. We discuss this further, after providing an exposition of the game rules.

\begin{figure}[t]
  \centering
  \begin{subfigure}[b]{0.32\textwidth}
    \centering
    \includegraphics[width=\linewidth]{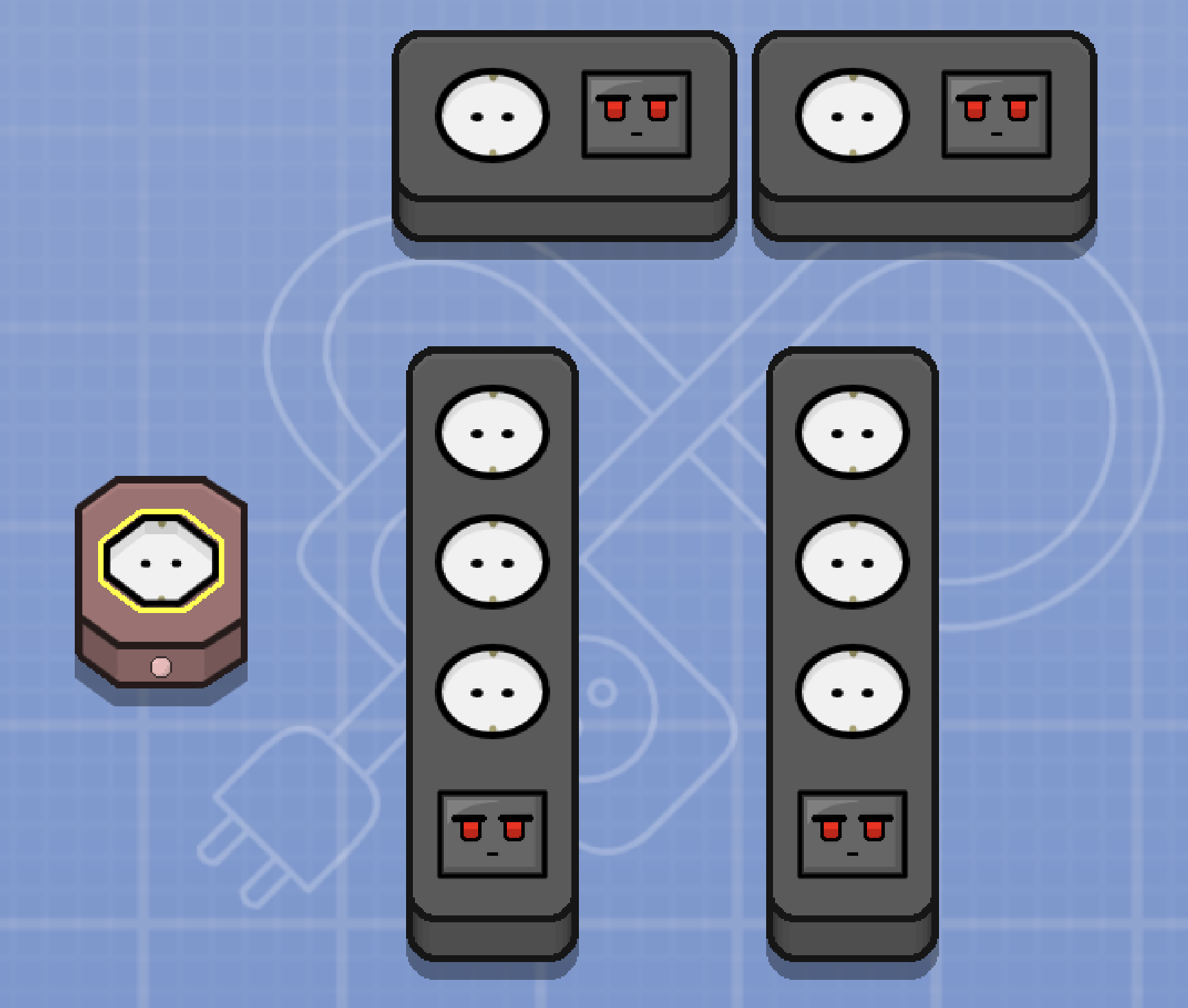}
    \caption{}\label{fig:play-a}
  \end{subfigure}\hfill
  \begin{subfigure}[b]{0.32\textwidth}
    \centering
    \includegraphics[width=\linewidth]{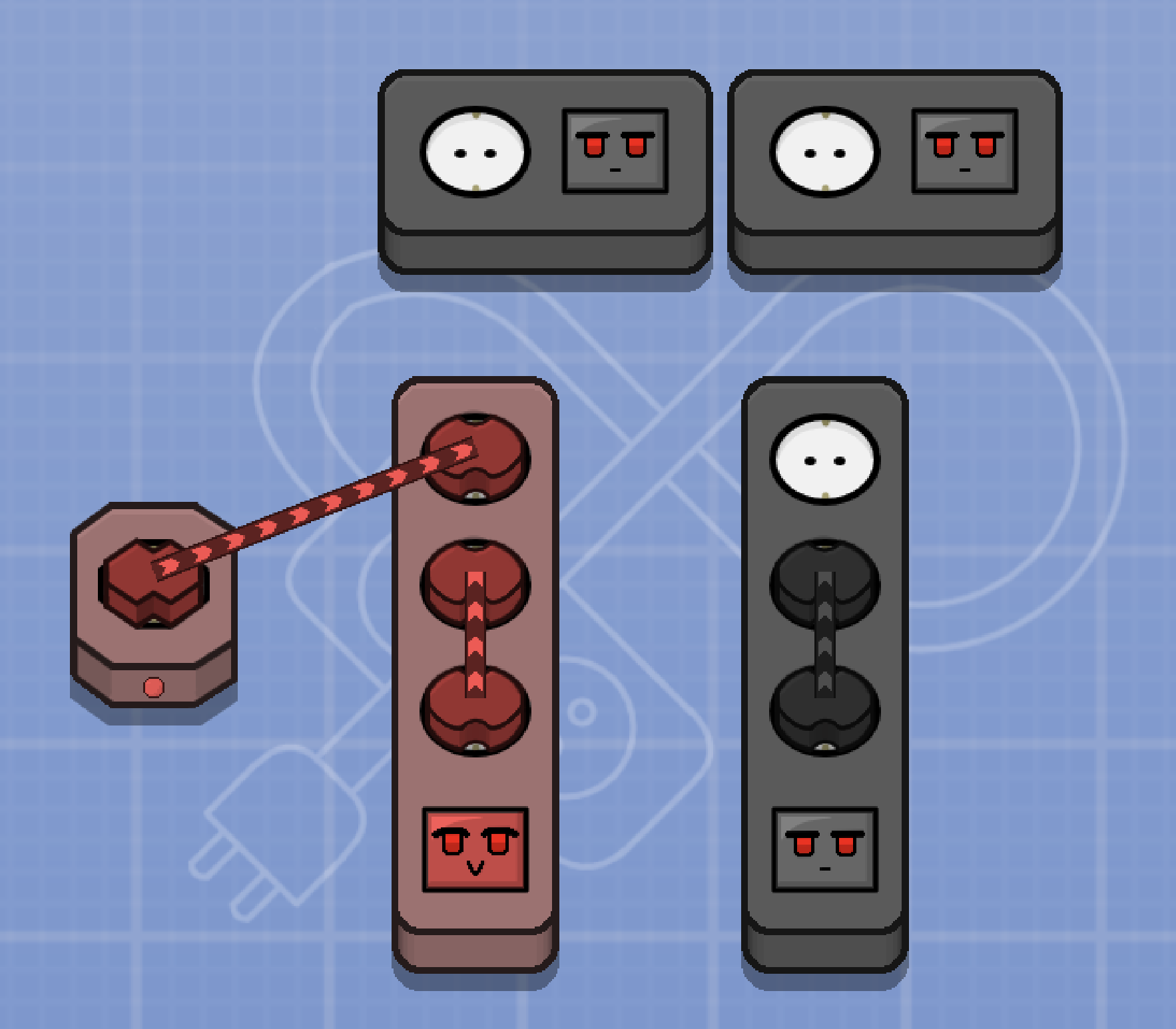}
    \caption{}\label{fig:play-b}
  \end{subfigure}\hfill
  \begin{subfigure}[b]{0.32\textwidth}
    \centering
    \includegraphics[width=\linewidth]{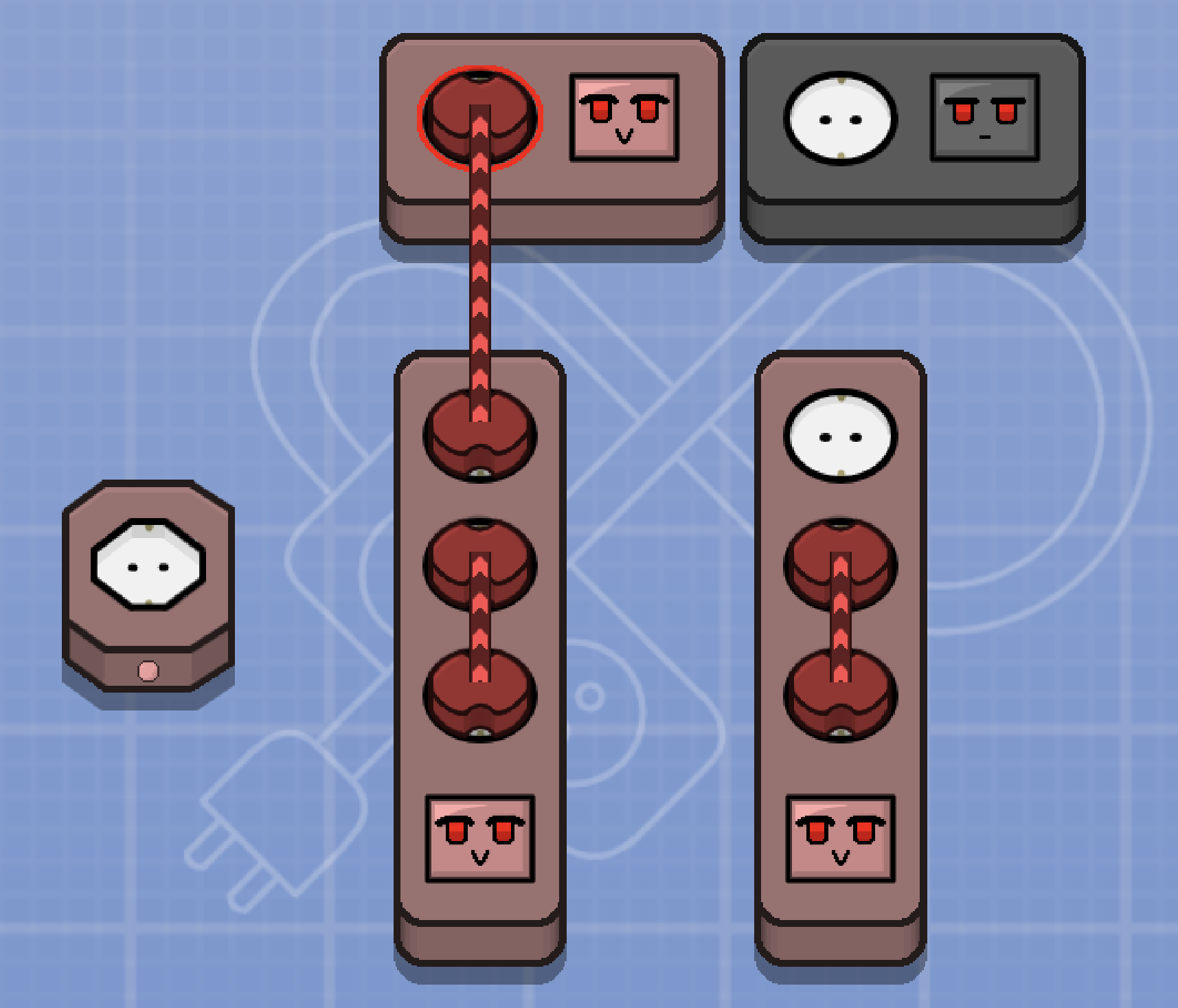}
    \caption{}\label{fig:play-c}
  \end{subfigure}
  \caption{A play of \IE{} on an instance with one red power source (the
  octagon on the left) and four boards, each carrying a red color requirement
  shown by the red eyes of its face. (a) The initial configuration: every
  board is unenergized and the source is detached. (b) The source is attached
  and emits its color along a wire into the upper socket of the left vertical
  board. Two of its three sockets are joined into a directed cycle (the third
  stays free to pass energy on), and the board becomes energized red so its
  face smiles. The right vertical board is wired into a cycle of the same
  shape but remains unenergized, drawn cold, since no energy has yet entered
  it. (c) The source is detached. The charged cycle continues to circulate
  energy without it, and because sockets sharing a board are joined for free,
  the energy propagates along cross-board wires to the remaining boards. Every
  board now displays its requirement met, and the instance is won with no
  source attached. The sustaining directed cycle from (b) is drawn
  schematically in Fig.~\ref{fig:loops}.}
  \label{fig:play}
\end{figure}

An instance is played on a collection of \emph{boards} (the set $B$), each
holding one or more circular \emph{sockets} (collectively $P$) and displaying a
cartoon face that smiles once the board is satisfied. Energy is supplied by
external \emph{power sources} (the set $S$), drawn as octagonal units, and
travels along directed \emph{wires} that
the player draws between sockets. In the full game a board also
carries a \emph{color} requirement, shown by the eyes of its face. Color is a
separate constraint that we isolate in Appendix~\ref{app:color} and ignore until
then, so the main text studies the colorless game. Two rules govern the flow
of energy:
\begin{enumerate}
  \item The sockets on a board are \emph{intrinsically connected}, so
  energizing any one of them energizes the whole board and makes the
  remaining sockets available as outgoing sources, and
  \item each socket admits at most one wire, so that a socket either
  transmits energy or receives it, but never both.
\end{enumerate}

The object of the game is to keep the boards powered \emph{after every source
has been disconnected}, which compels the player to route energy into directed
cycles that circulate without external input. Fig.~\ref{fig:play} follows a
single play through this mechanism, and Fig.~\ref{fig:unconstrained}
(Section~\ref{sec:unconstrained}) contrasts a small winnable instance with one
that cannot be won.

The step from Fig.~\ref{fig:play-b} to Fig.~\ref{fig:play-c} is where the
mechanism operates. While the source is present it charges the connected
structure and after it is removed the boards stay charged because the energy is
trapped in a self-sustaining cycle. The player must therefore reach a state in
which energy keeps propagating to every board after the source is gone.
Section~\ref{sec:formal} defines energization as the least fixed point of a
monotone propagation rule, i.e., an instance is won when its energized set remains a
nonempty fixed point after the sources are deleted. We encourage readers to
try the variants at \url{https://infinite-energy-peer-review.netlify.app}.

This requirement places \IE{} between two families of previously analyzed
games: the \emph{trajectory} games, such as platformers and sliding-block
puzzles, whose solutions are sequences of moves~\cite{aloupis,forisek,hearn2005},
and the \emph{consistency} puzzles, such as Minesweeper, whose solutions are
static labelings checked against local constraints~\cite{debondt,kaye}.

\IE{} shares with the trajectory games a winning condition defined over a
process rather than a state, so the final configuration does not certify a
win. The wire graph left behind once the sources are withdrawn is a union of
paths and cycles, and it does not record whether its loops were ever charged.
In Fig.~\ref{fig:play-b} the un-energized cycle on the right is structurally
identical to the charged one on the left, so the terminal graph
underdetermines the energized set. A frozen frame of generalized Super Mario
Bros.\ is uninformative in the same way, certifying nothing about the legality
of the play that produced it~\cite{aloupis}. The two families diverge in the
certificate their processes admit. For a trajectory game the certificate is
irreducibly a sequence of moves, checked by forward simulation, and
generalized instances need not admit any short certificate, which places the
reconfiguration framework underlying many of them in
\PSPACE~\cite{hearn2004,hearnbook}. For \IE{} one might hope that charging is
monotone, so a static witness suffices: the final loops, each with a route in
the connectability graph from a source to a free socket on one of its boards.
This hope fails. Some instances are winnable only by charging a temporary
\emph{battery}, a loop built and then broken to free its sockets
(Proposition~\ref{prop:battery}), so the final loops do not on their own
certify a win. They must also be chargeable in some order, with batteries
where needed and without the charging routes competing for socket capacity,
and the final wiring has erased that schedule. We therefore take the
certificate to be a short winning play, and reduce the existence of one to a
single conjecture: that a winning play never needs to charge any socket twice
(Section~\ref{sec:upper}). Without this conjecture, the problem is \NP-hard and in
\PSPACE.

The game layers two further constraints on this mechanic. Wires have bounded
length, so not every pair of sockets may be joined. In the standard game this
restriction comes from a maximum wire length, and only sockets closer than
that length can be connected. We relax it by taking connectability to be
arbitrary, given by a connectability matrix $M$ supplied with the instance,
and we return to the geometric version in Section~\ref{sec:geometric}.

The second restriction assigns a color to each source and a color requirement
to each board, with colors mixing additively as energy flows. We describe only
the colorless game and defer the colored variant to
Appendix~\ref{app:color}. Since the game is \NP-hard without color, the
colored variant is \NP-hard too, but a matching upper bound remains open.

\begin{table}[t]
  \centering
  \begin{tabular}{lll}
    \toprule
    \textbf{Variant} & \textbf{Complexity} & \textbf{Reference} \\
    \midrule
    Unrestricted $M$ (all-ones) & \TC-complete & Theorem~\ref{thm:tc0} \\
    Arbitrary $M$ & \NP-hard, in \PSPACE & Theorems~\ref{thm:nphard}, \ref{thm:pspace} \\
    Geometric, $\R^{n-1}$ & \NP-hard, in \PSPACE & Corollary~\ref{cor:geom} \\
    Geometric, $\R^k$, $k \ge 2$ & \NP-hard, in \PSPACE & Theorem~\ref{thm:fixeddim} \\
    Geometric, $\R^1$ & open & --- \\
    Colored & \NP-hard & Appendix~\ref{app:color} \\
    \bottomrule
  \end{tabular}
  \caption{Complexity of \IE{} across variants. For the \NP-hard rows,
  \NP-completeness would follow from Conjecture~\ref{conj:rf} (recharge-free
  plays suffice), which we state and support but do not prove.}
  \label{tab:results}
\end{table}

\paragraph{Our results.} We separate the combinatorics of the game from its
geometry. When connectability is unrestricted ($M$ all-ones), winnability
reduces to two arithmetic conditions: at least $2|B|$ sockets and some board
with at least three sockets (Theorem~\ref{thm:char}), and the induced
predicate is \TC-complete, hence in \cL{} (Theorems~\ref{thm:L},
\ref{thm:tc0}, Proposition~\ref{prop:tc0}). With $M$ given arbitrarily the
problem becomes \NP-hard by reduction from \HP{} (Theorem~\ref{thm:nphard}),
and lies in \PSPACE{} (Theorem~\ref{thm:pspace}). We reduce \NP-membership to
a conjecture that recharge-free plays suffice (Conjecture~\ref{conj:rf},
Section~\ref{sec:upper}), which would give \NP-completeness. This hardness is
not an artifact of unrealizable matrices. Every symmetric $M$ is realizable as
a distance-threshold graph in $\R^{n-1}$, where $n = |P| + |S|$
(Theorem~\ref{thm:realize}), so the problem stays \NP-hard under that
geometric restriction (Corollary~\ref{cor:geom}). In fixed dimension it
remains \NP-hard for every $k \ge 2$ (Theorem~\ref{thm:fixeddim}). The
one-dimensional case is open. Finally, the colored variant is \NP-hard, though
we do not settle a matching upper bound (Appendix~\ref{app:color}).
Table~\ref{tab:results} collects these results.

\subsection{Graph-Theoretic Formalization.}\label{sec:formal}

\begin{definition}[Instance]\label{def:instance}
A \textup{(colorless)} game instance is a tuple $\mathcal{I} = (P, S, B, M)$
where
\begin{itemize}
  \item $P$ is a finite set of \textup{sockets}.
  \item $S$ is a finite set of \textup{sources}, with $P \cap S = \emptyset$.
  \item $B$ is a partition of $P$ into nonempty \textup{boards}, with
  $B \neq \emptyset$ (the empty instance wins vacuously and is excluded). We
  write $b(p)$ for the board containing socket $p$.
  \item $M \in \{0,1\}^{(P \cup S) \times (P \cup S)}$ is a symmetric
  \textup{connectability matrix}, where $M[x,y] = 1$ means a wire may join $x$
  and $y$. We set $M[s,s'] = 0$ for all $s, s' \in S$: a wire between two
  sources could never carry energy (a source only supplies, never receives, by
  Definition~\ref{def:config}), so forbidding such wires loses no plays and
  lets us treat every socket uniformly.
\end{itemize}
\end{definition}

A wire carries energy in one direction, so we model a wiring as a directed
graph.

\begin{figure}[t]
  \centering
  \hspace*{-1.6pt}\begin{tikzpicture}[x=1pt,y=-1pt]
    \draw[board] (137,126.1) rectangle (193.7,154.5);
    \draw[sock] (148.35,140.25) circle (4.25pt);
    \draw[sock] (182.35,140.25) circle (4.25pt);
    \draw[wire] (150.6,136.3) .. controls (156.3,126.4) and (174.3,126.4) .. (179.6,135.5);
    \draw[impl] (180.1,144.2) .. controls (174.3,154.2) and (156.3,154.2) .. (151.1,145.0);
    \node[font=\scriptsize] at (148.3,151.8) {$p_1$};
    \node[font=\scriptsize] at (182.3,151.8) {$p_2$};
    \node[font=\scriptsize, align=center] at (165.4,170.5) {2-cycle,\\one board};
    \draw[board] (247.5,116.2) rectangle (275.9,164.4);
    \draw[board] (309.9,116.2) rectangle (338.2,164.4);
    \draw[sock] (261.75,127.55) circle (4.25pt);
    \draw[sock] (261.75,153.05) circle (4.25pt);
    \draw[sock] (324.05,127.55) circle (4.25pt);
    \draw[sock] (324.05,153.05) circle (4.25pt);
    \draw[wire] (266.3,127.5) -- (318.6,127.5);
    \draw[impl] (328.0,129.8) .. controls (335.1,133.9) and (335.1,146.7) .. (330.9,150.0);
    \draw[wire] (319.5,153.0) -- (267.2,153.0);
    \draw[impl] (257.8,150.8) .. controls (250.7,146.7) and (250.7,133.9) .. (256.9,130.3);
    \node[font=\scriptsize, align=center] at (292.9,170.5) {4-cycle,\\two boards};
    \draw[wire] (380.8,128.9) -- (399.7,128.9);
    \node[font=\scriptsize, anchor=west, inner sep=0pt] at (408,128.6) {wire (in $E$)};
    \draw[impl] (380.8,143.1) -- (399.7,143.1);
    \node[font=\scriptsize, anchor=west, inner sep=0pt] at (408,142.6) {implicit board edge};
  \end{tikzpicture}
  \caption{The two smallest loops (Definition~\ref{def:loop}), drawn as
  directed cycles. Solid arrows are wires of $E$. Dashed arrows are implicit
  board edges of $E_B$, which cost no wire slot. Edge types alternate, so a
  loop of length $2k$ uses $k$ wires and $k$ implicit edges. The 2-cycle is
  the loop drawn as the red wires with both ends on the same board in Fig.~\ref{fig:play-c}.}
  \label{fig:loops}
\end{figure}
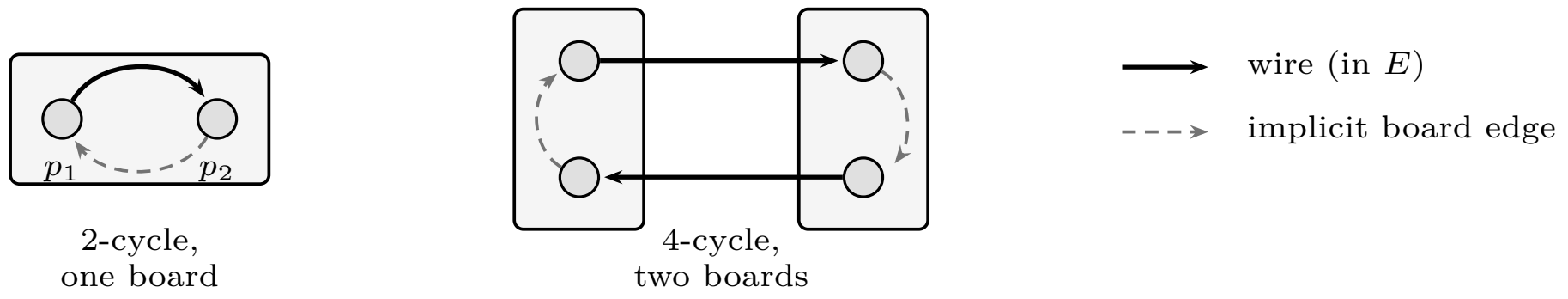

\begin{definition}[Configuration]\label{def:config}
A \textup{configuration} for $\mathcal{I}$ is a directed graph
$W = (P \cup S, E)$, where each edge is a wire oriented from the
energy-supplying end to the receiving end, satisfying (with $\deg^+(v)$ and
$\deg^-(v)$ being the out- and in-degree of $v$ in $W$):
\begin{enumerate}
  \item \textbf{One wire per part:} $\deg^+(v) + \deg^-(v) \le 1$ for every
  $v \in P \cup S$.
  \item \textbf{Connectable:} $(x,y) \in E \implies M[x,y] = 1$.
  \item \textbf{Sources only supply:} $\deg^-(s) = 0$ for every $s \in S$ (no
  wire points into a source).
\end{enumerate}
\end{definition}
\noindent Since the vertex set $P \cup S$ is fixed by $\mathcal{I}$, a configuration is
determined by its wire set $E$. We write the configuration as $E$ when only
the wires matter.

Sockets on the same board pass energy without a wire. We record this with
edges that do not count against the degree bound.

\begin{definition}[Implicit edges and completion graph]\label{def:implicit}
The set of \textup{implicit edges} is
\[
  E_B = \bigl\{ (p,q), (q,p) \bigm| p \neq q,\; b(p) = b(q) \bigr\},
\]
(energy flows either way within a board). The \textup{completion graph} of a
configuration $W = (P \cup S, E)$ is $G' = (P \cup S, E \cup E_B)$.
\end{definition}

\begin{definition}[Loop]\label{def:loop}
A \textup{loop} is a directed cycle in $G'$ whose edges alternate between
wires (from $E$) and implicit edges (from $E_B$):
\[
  p_1 \xrightarrow{\text{wire}} p_2 \xrightarrow{\text{impl.}} p_3
  \xrightarrow{\text{wire}} p_4 \xrightarrow{\text{impl.}} \cdots
  \xrightarrow{\text{impl.}} p_1.
\]
\end{definition}
\noindent Two facts follow immediately from Definition~\ref{def:loop}. First, because
the edge types alternate around the cycle, every loop has \emph{even} length
$2k$, using exactly $k$ wires and $k$ implicit edges. We call these $k$ wires
the \emph{$L$-edges} of the loop $L$. Each socket of $L$ carries exactly one of
them. Second, every socket on a loop has one incoming and one outgoing cycle
edge of which exactly one is a wire, so a loop never exceeds the
one-wire-per-socket budget.

It remains to say how energy spreads and what it means to win.

\begin{definition}[Energization]\label{def:energ}
Fix a configuration $W$. A socket is \textup{energized} in $W$ if, in the
completion graph $G'$, it is reachable from some energized socket or a power
source by a directed path of wires and implicit edges. A board is
\textup{energized} if at least one of its sockets is energized (equivalently, because of the
implicit edges, if one socket is energized all the sockets on the board are). A loop is said to be \textup{charged} in $W$
if every socket on it is energized in $W$.
\end{definition}

The current wiring does not by itself determine energization: a charged loop
keeps its sockets energized even after the source that charged it is
unplugged (Fig.~\ref{fig:loops}). Energization is therefore relative to a
play's history, not to the current wiring alone. Appendix~\ref{app:membership}
makes this precise with a recurrence $\en_t$ over the configurations a play
traverses. Informally, a socket is energized at step $t$ when it is reachable,
through wires and board edges, from a source or from a loop that was charged
earlier and whose wires still stand. This persistence is the source of the
problem's difficulty.

\begin{definition}[Move and play]\label{def:move}
A \textup{move} is a pair $m = (e, b)$ where $e = (p,q)$ is a directed edge
with $p, q \in P \cup S$ and $b \in \{\textsf{connect}, \textsf{disconnect}\}$.
Applied to a configuration $E$, the move yields
\[
  E \cdot m =
  \begin{cases}
    E \cup \{e\}, & b = \textsf{connect},\\
    E \setminus \{e\}, & b = \textsf{disconnect},
  \end{cases}
\]
and is \textup{legal} at $E$ if $E \cdot m$ is a valid configuration
(Definition~\ref{def:config}) and, additionally, $e \notin E$ when
$b = \textsf{connect}$ and $e \in E$ when $b = \textsf{disconnect}$.

A \textup{play} is a finite sequence of moves $m_1, \dots, m_T$ such that,
writing $E_0 = \emptyset$ and $E_t = E_{t-1} \cdot m_t$, each $m_t$ is legal
at $E_{t-1}$. We call $E_0, \dots, E_T$ the \textup{configurations traversed by
the play}.
\end{definition}

\begin{ieproblem}[Infinite Energy]
Given an instance $\mathcal{I}$, is there a play whose final configuration
$E_T$ connects no source yet leaves every board energized at $T$
\textup{(Appendix~\ref{app:membership})}?
\end{ieproblem}

\section{Unconstrained Infinite Energy is in \cP}\label{sec:unconstrained}

The \emph{unconstrained} variant \UIE{} is the special case in which the
connectability matrix imposes no restriction: $M[x,y] = 1$ for every
socket--socket and source--socket pair (only source--source wires remain
forbidden). The input is then only $(P, S, B)$ with $S \neq \emptyset$, and
we ask whether the instance is winnable, and we give a linear time algorithm.

\begin{theorem}\label{thm:char}
An unconstrained instance $(P, S, B)$ with $S \neq \emptyset$ is winnable if
and only if
{\setlength{\leftmargini}{23.8pt}%
\renewcommand{\labelenumi}{(\roman{enumi})}%
\begin{enumerate}
  \item $|P| \ge 2|B|$, and
  \item some board has at least $3$ sockets.
\end{enumerate}}
\end{theorem}

\begin{figure}[t]
  \centering
  \begin{tikzpicture}[x=1pt,y=-1pt]
    \draw[src] (168.5,157.4) -- (175.2,157.4) -- (179.9,162.1) -- (179.9,168.9)
      -- (175.2,173.6) -- (168.5,173.6) -- (163.7,168.9) -- (163.7,162.1) -- cycle;
    \node[font=\scriptsize, text=black!55] at (171.8,148.6) {(removed)};
    \node[font=\scriptsize] at (171.8,182.0) {$s$};
    \draw[board] (200.2,130.1) rectangle (228.5,200.9);
    \node[font=\scriptsize] at (214.4,121.5) {$b^*$};
    \draw[sock] (214.35,142.85) circle (4.25pt);
    \draw[sock] (214.35,165.55) circle (4.25pt);
    \draw[sock] (214.35,188.15) circle (4.25pt);
    \draw[board] (268.2,141.4) rectangle (296.6,189.6);
    \draw[sock] (282.35,152.75) circle (4.25pt);
    \draw[sock] (282.35,178.25) circle (4.25pt);
    \draw[wire] (218.3,167.8) .. controls (224.4,171.3) and (224.4,182.4) .. (221.2,185.1);
    \draw[impl] (210.4,185.9) .. controls (204.3,182.4) and (204.3,171.3) .. (209.6,168.3);
    \draw[wire] (218.9,142.8) .. controls (242.2,142.8) and (254.5,152.8) .. (276.9,152.8);
    \node[font=\scriptsize] at (240.2,213.4) {(a) winnable};
    \draw[src] (352.7,157.4) -- (359.4,157.4) -- (364.2,162.1) -- (364.2,168.9)
      -- (359.4,173.6) -- (352.7,173.6) -- (348.0,168.9) -- (348.0,162.1) -- cycle;
    \node[font=\scriptsize] at (356.1,182.0) {$s$};
    \draw[board] (378.8,141.4) rectangle (407.1,189.6);
    \draw[wsock] (392.95,152.75) circle (4.25pt);
    \draw[wsock] (392.95,178.25) circle (4.25pt);
    \draw[board] (435.5,141.4) rectangle (463.8,189.6);
    \draw[wsock] (449.65,152.75) circle (4.25pt);
    \draw[wsock] (449.65,178.25) circle (4.25pt);
    \node[font=\scriptsize] at (421.1,213.4) {(b) not winnable};
  \end{tikzpicture}
  \caption{Two unconstrained instances (final configurations. No source
  attached. Shaded sockets are energized). (a) A winnable instance: $b^*$ has
  three sockets, so two of them form a self-sustaining 2-cycle (charged
  earlier through $s$, then $s$ removed) and the third exports energy to the
  other board. (b) An unwinnable instance: two boards of two sockets each.
  Here $|P| = 2|B|$, but no board has three sockets, so any loop would occupy
  both sockets of each board it meets and leave none off the loop to charge it
  (Lemma~\ref{lem:three}). Once $s$ is removed nothing stays energized.}
  \label{fig:unconstrained}
\end{figure}
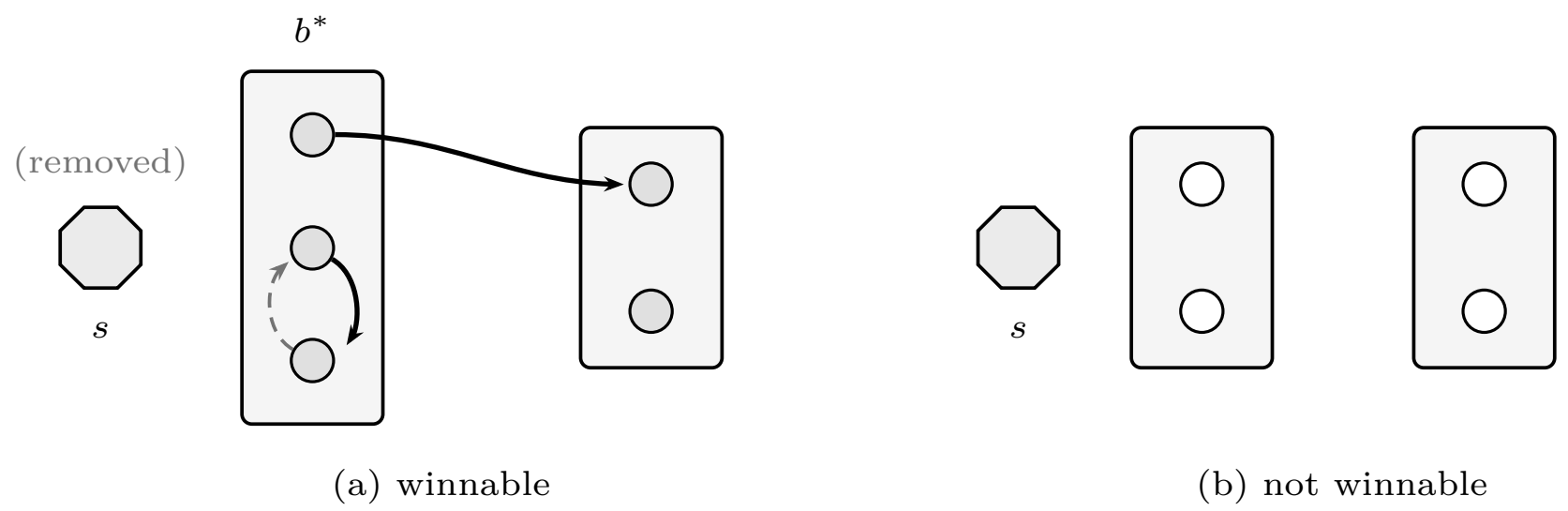

\noindent Both conditions are checkable in $O(|B|)$ time, so \UIE{}
$\in \cP$. Fig.~\ref{fig:unconstrained} shows a small winnable and a small
unwinnable instance. The construction behind sufficiency is illustrated in
Fig.~\ref{fig:sufficiency} (Appendix~\ref{app:proofs}).

\subsection{Sufficiency}

\begin{proof}[sketch]
Fix a board $b^*$ with $|b^*| \ge 3$ and bootstrap a self-sustaining 2-cycle on
two of its sockets (charge through a source, then disconnect it). Maintain a
pool $\mathcal{F}$ of free energized sockets. Energizing a board $b$ nets
$|b| - 2$ to $\mathcal{F}$, so processing boards in non-increasing size keeps
$\mathcal{F}$ nonempty exactly when $\sum_b (|b| - 2) = |P| - 2|B| \ge 0$. Full
proof in Appendix~\ref{app:proofs}.
\end{proof}

\subsection{Necessity}

Both necessity proofs start from a winning final configuration $W$, which by
definition has no source connected yet every board energized. Since the base
case of energization is empty without sources, all energy in $W$ originates
from loops.

\begin{lemma}\label{lem:three}
If every board has at most $2$ sockets, the instance is not winnable.
\end{lemma}

\begin{proof}[sketch]
A charged loop's sockets are all wire-saturated, so it can have been charged
only through an implicit edge from an off-loop socket on a board it meets. Because implicit edges only exist within the same board, that board then has at least three sockets. Full proof in
Appendix~\ref{app:proofs}.
\end{proof}

\begin{lemma}\label{lem:count}
If $|P| < 2|B|$, the instance is not winnable.
\end{lemma}

\begin{proof}[sketch]
Contract each board to a node. Every energized board has in-degree at least
one, so there are at least $|B|$ socket--socket wires and each uses two of the
$|P|$ socket slots, so $|P| \ge 2|B|$. Full proof in
Appendix~\ref{app:proofs}.
\end{proof}

Together, Lemmas~\ref{lem:three} and~\ref{lem:count} give the
necessary direction of Theorem~\ref{thm:char}.

\section{Hardness Lower Bound for Unconstrained Infinite Energy}

Theorem~\ref{thm:char} reduces \UIE{} to checking two arithmetic conditions on
$(P, B)$: that $|P| \ge 2|B|$ and that some board has at least three sockets.
We pin down the exact complexity of this predicate, which turns out to be far
below \cP.

Membership in \cP{} already shows the unconstrained game is easy, but not
\emph{how} easy. The size test $|P| \ge 2|B|$ is a comparison of two counts, and
we show below that this counting is not only sufficient but necessary. We use
the standard circuit classes (see~\cite{arora,vollmer}). \AC{} is the class of
languages decided by families of constant-depth, polynomial-size Boolean
circuits over unbounded fan-in $\wedge, \vee, \neg$ gates. \TC{} is the same
with unbounded fan-in \emph{threshold} (equivalently majority) gates also
allowed. Hence $\mAC \subseteq \mTC \subseteq \mathsf{L} \subseteq \mathsf{P}$,
and \MAJ{} $\notin \mAC$~\cite{arora,vollmer}, so the inclusion
$\mAC \subseteq \mTC$ is strict. Locating the predicate in \TC{} but outside
\AC{} (under the encodings, uniformity, and reductions fixed below) is the
sense in which its only difficulty is counting.

\begin{theorem}\label{thm:L}
\UIE{} $\in \mathsf{L}$.
\end{theorem}

\begin{proof}
Scan the input once, maintaining three quantities: a running count of boards
($|B|$), a running count of sockets ($|P| = \sum_b |b|$), and a single bit
recording whether any board seen so far has more than two sockets. At the end,
accept iff the bit is set and $|P| \ge 2|B|$. Each counter holds a number at
most the input size $n$, so uses $O(\log n)$ bits. The comparison
$|P| \ge 2|B|$ is computable in the same space. Hence the procedure runs in
deterministic logspace.
\end{proof}

The predicate is also constant-depth: $|P| \ge 2|B|$ is one threshold over the
input, and ``some board exceeds two sockets'' is a disjunction of local
checks. This gives us the following proposition:

\begin{proposition}\label{prop:tc0}
\UIE{} $\in \mTC$.
\end{proposition}
\noindent We complement this with a matching lower bound: the problem is as hard as
counting, so it cannot lie in \AC~\cite{arora,vollmer}.

\begin{theorem}\label{thm:tc0}
\MAJ{} reduces to \UIE{} by a projection, in particular by a many-one \AC{}
reduction. Since \MAJ{} is \TC-complete under \AC{} Turing
reductions~\cite{vollmer}, \UIE{} is \TC-complete under \AC{} Turing
reductions.
\end{theorem}

\begin{proof}[sketch]
Map $x \in \{0,1\}^m$ to one board per bit: a three-socket board when
$x_i = 1$ and a singleton when $x_i = 0$. Then $|B| = m$ and $|P| = m + 2s$
with $s = \sum_i x_i$, so $|P| \ge 2|B| \iff s \ge m/2$, and a 1-bit supplies
the required three-socket board. Thus the instance accepts iff $x$ is
majority-1. Each output board depends on one input bit, so the map is a
projection, and a projection is an \AC{} reduction. Full proof in
Appendix~\ref{app:proofs}.
\end{proof}
\noindent Two conventions make the completeness statement precise. We take the standard
encoding in which the $m$ boards are listed in input order in fixed-width
records, so the $i$-th board is the one-bit projection above (later output
positions do not depend on earlier bits), and we read \TC{} and \AC{} as
\textsc{dlogtime}-uniform, under which \MAJ{} is complete and
$\mTC \subseteq \mathsf{L}$ hold. \MAJ{} accepts on ties (at least $m/2$ ones),
matching the size test. A
projection gives only many-one \MAJ-hardness, which already places \UIE{}
outside \AC, whereas the preceeding completeness uses that \MAJ{} is \TC-complete
under \AC{} \emph{Turing} reductions. Summarizing, \UIE{} sits at
$\mTC \subseteq \mathsf{L} \subseteq \mathsf{P}$, a sharp drop from the
\NP-hardness of the full game.

\section{Infinite Energy is \NP-Hard}\label{sec:nphard}

We reduce from \HP.

\begin{definition}[Hamiltonian Path]
Given an undirected graph $G = (V, E)$, decide whether $G$ has a
\textup{Hamiltonian path}: an ordering $v_{\sigma(1)}, \dots, v_{\sigma(n)}$ of
all $n = |V|$ vertices such that $\{v_{\sigma(k)}, v_{\sigma(k+1)}\} \in E$ for
every $k < n$.
\end{definition}

\noindent \HP{} is \NP-complete~\cite{gareyjohnson}.

\begin{theorem}\label{thm:nphard}
\IE{} is \NP-hard.
\end{theorem}

\begin{figure}[t]
  \centering
  \includegraphics[width=\textwidth]{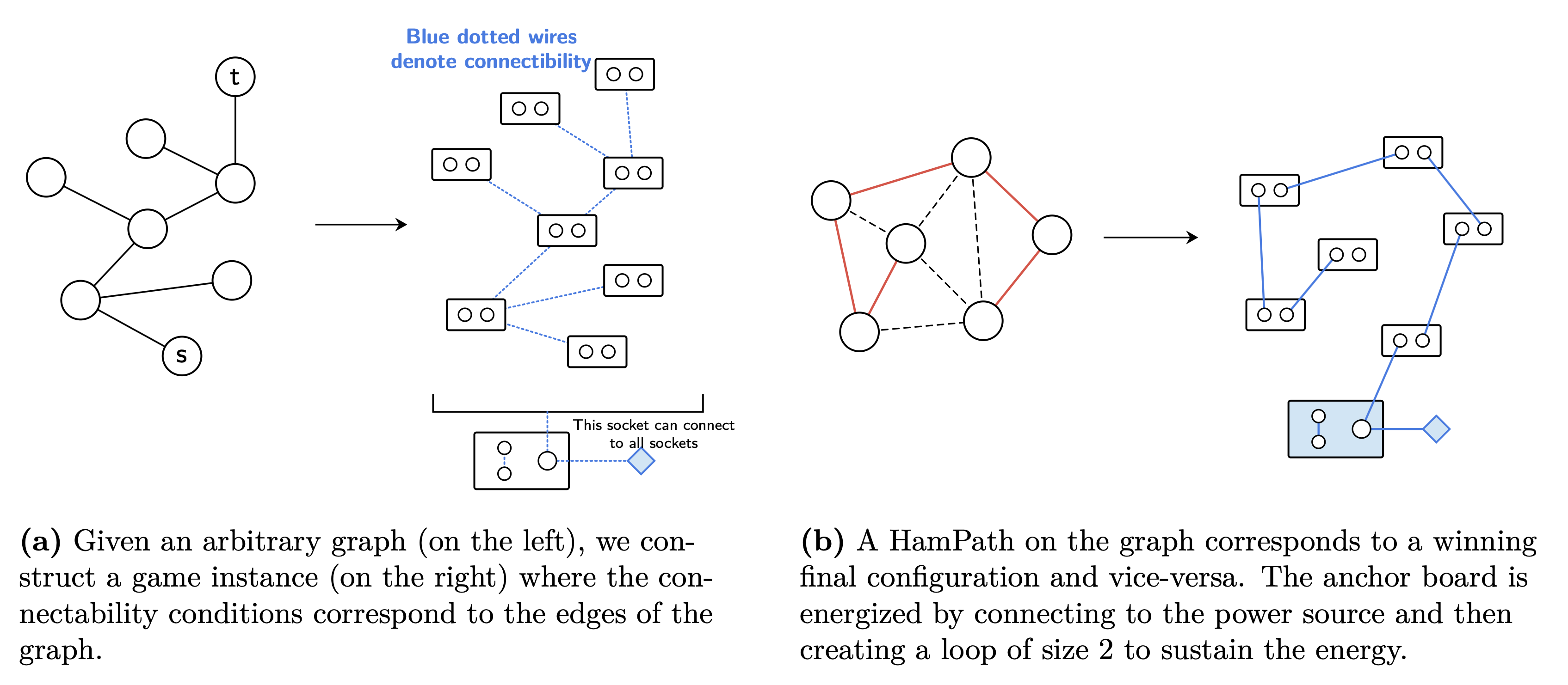}
  \caption{The reduction from Hamiltonian Path.}
  \label{fig:reduction}
\end{figure}

\paragraph{Construction} Given $G = (V, E)$ with $V = \{v_1, \dots, v_n\}$,
build an instance $\mathcal{I} = (P, S, B, M)$ as follows.
\begin{itemize}
  \item \emph{Node boards.} For each $v_i$ a board $b_i = \{p_i^L, p_i^R\}$
  with two sockets.
  \item \emph{Anchor board.} One board $b^* = \{p^*_1, p^*_2, p^*_3\}$ with
  three sockets.
  \item \emph{Source.} A single source $s$, so $S = \{s\}$.
\end{itemize}
Thus $|P| = 2n{+}3$ and $B = \{b_1, \dots, b_n, b^*\}$. The connectability
matrix $M$ permits exactly the following wires (and their symmetric entries):
{\renewcommand{\labelenumi}{(\alph{enumi})}%
\begin{enumerate}
  \item $p^*_1\wdash p^*_2$\ \quad (the two anchor sockets, only to each other).
  \item $s\wdash p^*_3$\quad and\ \quad $p^*_3\wdash p^X_i$ for every $i$ and
  $X \in \{L, R\}$.
  \item $p^X_i\wdash p^Y_j$\ \ whenever\ \ $\{v_i, v_j\} \in E$\ \ $(i \neq j,\ X, Y \in \{L, R\})$.
\end{enumerate}}
All other entries of $M$ are $0$. The construction is polynomial: $O(n)$
sockets and an $O(n^2)$ matrix filled directly from the adjacency relation of
$G$. See Fig.~\ref{fig:reduction}.

\paragraph{Correctness}

\begin{lemma}[Completeness]\label{lem:complete}
If $G$ has a Hamiltonian path, then $\mathcal{I}$ is winnable.
\end{lemma}

\begin{proof}[sketch]
Wire $s \to p^*_3$, form the anchor loop $p^*_1 \to p^*_2$, disconnect $s$,
then follow the Hamiltonian path, wiring $p^R_{\sigma(k)} \to
p^L_{\sigma(k+1)}$ board by board. Full proof in Appendix~\ref{app:proofs}.
\end{proof}

\begin{lemma}[Soundness]\label{lem:sound}
If $\mathcal{I}$ is winnable, then $G$ has a Hamiltonian path.
\end{lemma}

\begin{proof}[sketch]
Only $b^*$ has three sockets, so the sole charged loop is $\{p^*_1, p^*_2\}$.
Energy leaves $b^*$ only through $p^*_3$, and the degree-two node boards force
the energized region to be a single simple path visiting every one, i.e.\ a
Hamiltonian path. Full proof in Appendix~\ref{app:proofs}.
\end{proof}

\noindent Lemmas~\ref{lem:complete} and~\ref{lem:sound} prove
Theorem~\ref{thm:nphard}.

\section{Upper bounds: \PSPACE, and \NP{} modulo a conjecture}\label{sec:upper}

A win is defined by a \emph{play}, a sequence of moves, not by a static
picture: the same final wiring can be a win or a dead position, depending on
whether its loops were ever charged. A game state is captured by the pair
(current wiring, currently energized set), which takes polynomially many bits,
and the next state is a function of the state and the move
(Appendix~\ref{app:membership}). Deciding whether a winning state is reachable
from the empty wiring is therefore a reachability question over a state graph
whose states have polynomial size.

\begin{theorem}\label{thm:pspace}
\IE{} $\in \mathsf{PSPACE}$.
\end{theorem}

For \NP{} the certificate is a winning play, and two facts would suffice
(Appendix~\ref{app:membership}): a play is verifiable in time polynomial in
its length (Proposition~\ref{prop:verify}), and every winnable instance has a
short winning play (Theorem~\ref{thm:np}). The obstacle to the second is that
a win can require \emph{temporary batteries}, loops charged only to be broken
again. A seven-socket instance is winnable yet has no play that never breaks a
charged loop (Proposition~\ref{prop:battery}), so the natural ``charge and
keep'' strategy fails outright. However, no socket
ever needs its loop charged twice. Plays with this property, which we call
\emph{recharge-free}, have only $O(|P|)$ charged loops
(Lemma~\ref{lem:disjoint}), and we conjecture that one of polynomial length
always exists (Conjecture~\ref{conj:rf}).

\begin{theorem}
\IE{} is \NP-hard \textup{(Theorem~\ref{thm:nphard})} and lies in \PSPACE{}
\textup{(Theorem~\ref{thm:pspace})}. If Conjecture~\ref{conj:rf} holds, then
\IE{} is \NP-complete.
\end{theorem}

\section{Geometric Realizability of Connectability Matrices}\label{sec:geometric}

The connectability matrix abstracts a physical layout. In the real game two
sockets are joinable when they lie within a fixed wire length. We ask whether
requiring $M$ to come from points in $\R^k$ weakens the hardness of \IE. In
high dimension, geometry imposes no restriction, and hardness survives even in
fixed, small dimension.

\begin{definition}[$k$-realizable matrix]
Let $X = P \cup S$. A symmetric matrix $M \in \{0,1\}^{X \times X}$ with zero
diagonal is \textup{$k$-realizable} if there exist a map
$\varphi : X \to \R^k$ and a threshold $d > 0$ such that, for all distinct
$u, v \in X$,
\[
  M[u,v] = 1 \iff \|\varphi(u) - \varphi(v)\| \le d.
\]
Equivalently, $M$ is the adjacency matrix of a distance-threshold graph in
$\R^k$.
\end{definition}

\subsection{Universal realizability in \texorpdfstring{$\R^{n-1}$}{R\^{}(n-1)}}

\begin{theorem}\label{thm:realize}
Every symmetric $M \in \{0,1\}^{n \times n}$ with zero diagonal is
$(n-1)$-realizable.
\end{theorem}
\noindent The proof is in Appendix~\ref{app:realize}. The reduction of
Section~\ref{sec:nphard} produces an arbitrary connectability matrix, and
Theorem~\ref{thm:realize} places the sockets and sources of that matrix at
points of $\R^{n-1}$ whose pairwise distances put a wire exactly between the
pairs $M$ allows. The hardness of \IE{} therefore persists under this
geometric restriction.

\begin{corollary}\label{cor:geom}
\IE{} is \NP-hard, and \NP-complete if Conjecture~\ref{conj:rf} holds, even
when $M$ is required to be realizable in $\R^{n-1}$.
\end{corollary}

\subsection{Hardness in fixed dimension}

Corollary~\ref{cor:geom} leaves open whether hardness needs growing dimension.
It does not. The reduction of Section~\ref{sec:nphard} does not transfer
directly: there the anchor socket $p^*_3$ is connectable to \emph{every} node
socket, which is what lets a Hamiltonian path begin at any vertex. In a
fixed-dimensional realization a single socket cannot sit within the threshold
$d$ of arbitrarily many node boards spread across the lattice. We remove this
requirement by fixing the start of the path, so that the anchor need only
reach one node board.

\begin{definition}[Anchored Hamiltonian Path]
Given a graph $G = (V, E)$ and a vertex $r \in V$, decide whether $G$ has a
Hamiltonian path with $r$ as an endpoint.
\end{definition}

\begin{lemma}\label{lem:anchored}
\textsc{Anchored Hamiltonian Path} is \NP-complete even when $G$ comes with a
grid embedding in which the endpoint $r$ has degree one and is the strictly
highest vertex, so no other vertex lies in the row of $r$ or above it.
\end{lemma}

\begin{proof}[sketch]
Reduce from Hamiltonian path between two specified boundary terminals $s, t$
of a grid graph~\cite{itai}. Attach to $s$ an arm of new degree-two grid
vertices that leaves the bounding box through empty exterior cells and ends,
above the whole figure, in a degree-one tip $r$. Attach a similar arm to $t$
ending in $r'$. The arms force $r, r'$ to be the path's endpoints, and $r$ is
placed as the highest vertex. Full proof in Appendix~\ref{app:proofs}.
\end{proof}

\begin{theorem}\label{thm:fixeddim}
For every fixed $k \ge 2$, no-color \IE{} is \NP-hard \textup{(}and
\NP-complete if Conjecture~\ref{conj:rf} holds\textup{)} even when $M$ is
realizable in $\R^k$.
\end{theorem}

\begin{proof}[idea]
Place each grid vertex's two-socket board at its lattice point (sockets at
$(a \pm \varepsilon, b)$), so boards are connectable iff the vertices are
grid-adjacent, and put the anchor and source in the empty half-plane above the
graph, reaching only the degree-one terminal $r$. The reduction of
Section~\ref{sec:nphard} then applies with the path forced to start at $r$.
Full construction in Appendix~\ref{app:proofs}, with a worked $\R^2$ example
in Appendix~\ref{app:example}.
\end{proof}

\begin{credits}
\subsubsection{\ackname}
This work was supported by the Anusandhan National Research Foundation,
Government of India, under the Prime Minister Early Career Research Grant
ANRF/ECRG/2025/001136/ENS, and by the Vachani School of Advanced Computing,
Ashoka University under the Young Faculty Grant (00281).
\end{credits}

\bibliographystyle{alpha}
\bibliography{refs}

\clearpage
\appendix

\section{Membership: verification, \PSPACE, and recharge-free plays}\label{app:membership}

Fix an instance $\mathcal{I} = (P, S, B, M)$, with implicit edges $E_B$ and
completion graph $G'(E) = (P \cup S, E \cup E_B)$
(Definition~\ref{def:implicit}). A \emph{plug wire} joins two sockets. A
\emph{source wire} has its tail in $S$. A \emph{loop} is an alternating
directed cycle of $G'(E)$ (Definition~\ref{def:loop}), which uses at least one
plug wire.

\subsection{Energization is history-dependent}

A charged loop keeps its sockets energized after its source is gone, so which
sockets are energized depends on the play so far, not on the current wiring
alone. Fix a play $E_0 = \emptyset, \dots, E_T$ (Definition~\ref{def:move}).
Put $\en_0 = S$ and, for $t \ge 1$, let $\sus_t$ be the sockets lying on some
directed cycle of $G'(E_t)$ that uses a plug wire and all of whose sockets are
in $\en_{t-1}$. Let $\en_t$ be the least $X$ with
$S \cup \sus_t \subseteq X \subseteq P \cup S$ closed under flow ($u \in X$ and
$(u,v) \in E_t \cup E_B$ imply $v \in X$). A board is \emph{energized at $t$}
if it meets $\en_t$. The play \emph{wins at $T$} if $E_T$ carries no source
wire and every board is energized at $T$, i.e., the instance is \emph{winnable} if
some play wins.

This agrees with Definition~\ref{def:loop}. A directed cycle using a plug wire
and lying in $\en_{t-1}$ contains an alternating loop through that wire, so
$\sus_t$ is the flow-closure of the charged alternating loops' sockets. The
extra sockets it may name (co-board sockets carrying no wire) are energized
anyway, so $\en_t$ is unchanged. A loop, once all its sockets are in $\en$
while its wires stand, remains in $\sus$ thereafter, which is the
self-sustaining property of Fig.~\ref{fig:loops}.

\begin{proposition}\label{prop:verify}
Whether a play of length $T$ wins is decidable in time polynomial in $T$ and
$|\mathcal{I}|$.
\end{proposition}

\begin{proof}
Compute $\en_0, \dots, \en_T$ in order. Given $\en_{t-1}$, restrict $G'(E_t)$
to $\en_{t-1}$ and take its strongly connected components. A socket lies in
$\sus_t$ iff its component contains a plug wire, since then it sits on a
directed cycle through that wire inside $\en_{t-1}$. Then $\en_t$ is the set
reachable from $S \cup \sus_t$ in $G'(E_t)$, by one graph search. Each step
costs $O((|P| + |S|)^2)$, and there are $T$. Finally check that $E_T$ has no
source wire and every board meets $\en_T$.
\end{proof}

\subsection{A polynomial-space bound}

\begin{proof}[of Theorem~\ref{thm:pspace}]
A configuration is a wire set with each node on at most one wire, hence at
most $(|P| + |S|)/2$ plug wires and $|S|$ source wires, described in
$O((|P|+|S|)\log(|P|+|S|))$ bits. A \emph{state} is a pair $(E, X)$ with
$X = \en \subseteq P \cup S$, again of polynomial size. By \S A.1 the successor
of $(E, X)$ under a legal move $m$ is $(E \cdot m, \en')$, where $\en'$ is a
function of $X$ and $E \cdot m$ computable in polynomial time. Winning
states (no source wire, every board in $X$) are recognizable in polynomial
time. Deciding whether a winning state is reachable from $(\emptyset, S)$ in
this graph of $2^{\mathrm{poly}}$ states is in $\mathsf{NPSPACE} =
\mathsf{PSPACE}$.
\end{proof}

\subsection{A win can require temporary batteries}

A win may charge a loop only to break it later, using the loop as a
\emph{battery} that keeps a board energized while other sockets are rewired.
Batteries are not a convenience: some instances cannot be won without one.
Call a play \emph{monotone} if $\sus_0 \subseteq \sus_1 \subseteq \cdots$,
that is, if no charged loop is ever broken.

\begin{proposition}\label{prop:battery}
Some winnable instance has no monotone winning play.
\end{proposition}

\begin{proof}
Let $P = \{t\} \sqcup \{p, q, r\} \sqcup \{x, y, z\}$, with these three sets as
the boards, a single source $s$, and
\[
  M = \{\, xp,\ \ xz,\ \ pq,\ \ pt,\ \ yr,\ \ qr,\ \ zs,\ \ rs \,\}.
\]
The source reaches board $\{p,q,r\}$ only at $r$ and board $\{x,y,z\}$ only at
$z$. Writing $\overline{\,\cdot\,}$ for a disconnect, the play
\[
  s{\to}r,\ \  x{\to}z,\ \  p{\to}q,\ \  \overline{s{\to}r},\ \ 
  r{\to}y,\ \  \overline{r{\to}y},\ \  \overline{p{\to}q},\ \ 
  p{\to}t,\ \  y{\to}r
\]
wins: it lights $\{p,q,r\}$ from the source at $r$, charges the battery loop
$pq$, unplugs the source, relays energy through the freed socket $r$ by
$r{\to}y$ to charge the loop $xz$, dismantles the battery to free $p$,
installs the delivery $p{\to}t$, and relights $\{p,q,r\}$ from the now
self-sustaining board $\{x,y,z\}$ by $y{\to}r$. The final wiring
$\{xz, yr, pt\}$ carries no source wire and energizes all three boards.

No monotone play wins. To close $xz$, board $\{x,y,z\}$ must be energized with
$x, z$ free, so energy must enter at the spare $y$. The only wire at $y$ is
$yr$, so it arrives from $r$, and $r$ must be energized yet free to carry $yr$.
Board $\{p,q,r\}$ must therefore be lit while $r$ is free. The source meets
that board only at $r$, so it cannot light it without holding $r$, and
$\{x,y,z\}$, still uncharged at this point, cannot deliver into it. So the
board is lit by one of its own loops, and only $pq$ avoids $r$. Thus every win
charges the loop $pq$, which uses $p$. The only wire into board $\{t\}$ is
$pt$, so the win must also deliver through $p$. A monotone play never breaks
$pq$, so $p$ stays occupied and $pt$ can never be installed. Hence the
instance is winnable but not monotone-winnable, and the battery is
unavoidable.
\end{proof}

\subsection{Recharge-free plays and membership}

The winning play above breaks a loop but never rebuilds one: each socket
enters $\sus$ at most once. Call a play \emph{recharge-free} if for every
socket $v$ the set of steps with $v \in \sus_t$ is an interval, so that once a
loop through $v$ is broken, no loop through $v$ is ever charged again.
Recharge-freeness is strictly weaker than monotonicity: it allows the
discharges that Proposition~\ref{prop:battery} forces, but forbids charging
any socket twice.

\paragraph{The charging macro.} Let $R$ be a charged wiring we will not
disturb and $L$ a loop with a chosen \emph{closing wire} $e \in L$. A
\emph{support witness} for $L$ over $R$ is a simple directed path $Q$ with
$R \cup Q \cup (L \setminus \{e\})$ a valid wiring, both endpoints of $e$ free
in it, and $Q$ running from a source or a socket energized by $R$ to a socket
off $L$ but on a board of $L$.

\begin{lemma}[Charging macro]\label{lem:macro}
If $R$ is charged and $(Q, L, e)$ is a support witness over $R$, then $L$ can
be charged in $O(|P|)$ moves without disturbing $R$.
\end{lemma}

\begin{proof}
Add the wires of $Q$ and of $L \setminus \{e\}$. The witness makes this valid.
Energy from $Q$'s origin reaches its endpoint, and the arc $L \setminus \{e\}$
carries it around every board of $L$, so all their sockets are energized while
both endpoints of $e$ stay free. Add $e$: the cycle is present with all its
sockets energized at the previous step, so it becomes sustaining by the
recurrence of \S A.1. Remove $Q$. The wires of $L$ remain, so $L$ stays
charged. A simple path and a loop use $O(|P|)$ wires.
\end{proof}

\begin{lemma}\label{lem:disjoint}
In a recharge-free play the loops that ever become charged are pairwise
socket-disjoint. Hence at most $|P|/2$ loops are charged in all.
\end{lemma}

\begin{proof}
A socket carries at most one wire, so it lies on at most one loop at any step.
If two loops charged at different steps shared a socket $v$, then $v$ would
leave $\sus$ when the first loop broke and re-enter it when the second
charged, which recharge-freeness forbids. So the charged loops are pairwise
socket-disjoint, and each uses at least two sockets.
\end{proof}

\begin{conjecture}[Recharge-free suffices]\label{conj:rf}
Every winnable instance has a recharge-free winning play of length polynomial
in $|P|$.
\end{conjecture}

\begin{theorem}\label{thm:np}
If Conjecture~\ref{conj:rf} holds, then \IE{} $\in \mathsf{NP}$.
\end{theorem}

\begin{proof}
A play of polynomial length is verified in polynomial time
(Proposition~\ref{prop:verify}), so it is an \NP{} certificate.
\end{proof}

By Lemma~\ref{lem:disjoint} a recharge-free play charges only $O(|P|)$ loops,
each by an $O(|P|)$ macro (Lemma~\ref{lem:macro}). The content of
Conjecture~\ref{conj:rf} is that the routes and discharges between these
charges stay short, equivalently that a forced discharge never has to be
undone. Across all instances with $|P| \le 4$ and several million further
instances up to $|P| = 8$, every winnable instance had a recharge-free winning
play, and no shortest winning play ever recharged a socket. Whether some
family instead forces a recharge, and with it a superpolynomial minimal play,
is open. Such a family would place \IE{} above \NP{} and make the \PSPACE{}
bound of Theorem~\ref{thm:pspace} close to tight.

\section{Omitted proofs}\label{app:proofs}

\subsection*{Proof of Sufficiency in Theorem~\ref{thm:char}}

\begin{proof}[of sufficiency]
Fix a source $s$ and a board $b^*$ with $|b^*| \ge 3$, labelling its sockets
$p_1, p_2, p_3, \dots$

\smallskip\noindent\emph{Bootstrap.} Wire $s \to p_1$ (energizing $b^*$), then wire
$p_2 \to p_3$. The pair $\{p_2, p_3\}$ now forms a loop (wire $p_2 \to p_3$,
implicit edge $p_3 \to p_2$), and it is charged because $p_1$ is energized.
Disconnect $s$. The loop persists, so $b^*$ stays energized, and its remaining
$|b^*| - 2 \ge 1$ sockets are free and energized.

\begin{figure}[t]
  \centering
  \includegraphics{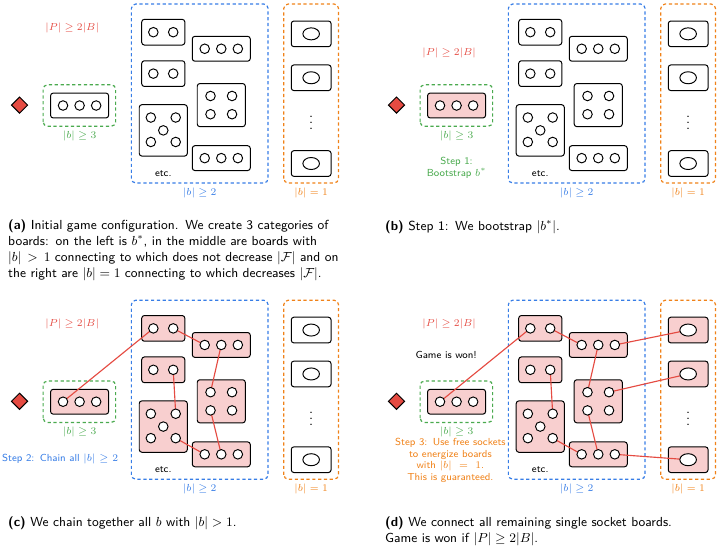}
  \caption{The construction behind sufficiency in Theorem~\ref{thm:char}. The
  boards fall into three groups: the single \emph{bootstrap board} $b^*$ (green, any
  board with at least three sockets), the boards with $|b| \ge 2$ (blue), and
  the singleton boards $|b| = 1$ (orange). Here $\mathcal{F}$ is the running
  pool of \emph{free energized sockets}. (a) The initial instance, with
  $|P| \ge 2|B|$. (b) Step 1 bootstraps a self-sustaining loop on $b^*$.
  (c) Step 2 chains every board with $|b| \ge 2$, each net adding
  $|b| - 2 \ge 0$ sockets to $\mathcal{F}$. (d) Step 3 spends one socket of
  $\mathcal{F}$ per singleton, which never runs out exactly when
  $|P| \ge 2|B|$.}
  \label{fig:sufficiency}
\end{figure}

\smallskip\noindent\emph{Chaining.} Maintain a pool $\mathcal{F}$ of free energized sockets,
initialized to the $|b^*| - 2$ leftover sockets of $b^*$. Process the
remaining boards in order of non-increasing size. To energize a board $b$,
draw a wire $f \to q$ from some $f \in \mathcal{F}$ to a socket $q \in b$. This
consumes $f$ and $q$ but energizes all of $b$, returning its other $|b| - 1$
sockets to $\mathcal{F}$. The net increase to $|\mathcal{F}|$ is
$-1 + |b| - 1 = |b| - 2$.

Processing larger boards first means every board with $|b| \ge 2$ (net change
$\ge 0$) is handled before any singleton (net change $-1$). When the first
singleton is processed the pool has size $\sum_{|b| \ge 2}(|b| - 2)$, and the
number of singletons is exactly $-\sum_{|b| = 1}(|b| - 2)$. Their difference is
$\sum_{b \in B}(|b| - 2) = |P| - 2|B| \ge 0$ by (i). Hence $\mathcal{F}$ never
empties prematurely, every board ends up energized, and no source is
connected.
\end{proof}

\subsection*{Proof of Lemma~\ref{lem:three}}

\begin{proof}
$W$ contains at least one loop $L$. Each socket of $L$ is incident to exactly
one wire and one implicit edge, and the implicit edge joins it to its
loop-neighbour on the \emph{same} board. Hence every board meeting $L$
contributes at least two sockets to it. The sockets of $L$ are all
wire-saturated, so for $L$ ever to have been charged, energy must have entered
it through an implicit edge from a socket \emph{not} on $L$, lying on a board
$b$ that also contains $L$-sockets. That board has at least two $L$-sockets
plus this extra one, so $|b| \ge 3$, contradicting the assumption.
\end{proof}

\subsection*{Proof of Lemma~\ref{lem:count}}

\begin{proof}
In $W$ all remaining wires are socket--socket. Let $w$ be their number. Each
wire uses two socket slots and each socket holds at most one, so
$w \le |P|/2$.

Contract each board to a single node, turning each wire into a directed edge
between board-nodes (an intra-board wire becomes a self-loop). Call the result
$\hat{G}$, with exactly $w$ edges. Energy enters a board only through a wire
into one of its sockets (implicit edges stay within a board), so every
energized board has in-degree at least $1$ in $\hat{G}$. As all $|B|$ boards
are energized, $w = \sum_b \deg^-_{\hat{G}}(b) \ge |B|$. Combining,
$|B| \le w \le |P|/2$, i.e.\ $|P| \ge 2|B|$.
\end{proof}

\subsection*{Proof of Lemma~\ref{lem:complete} (Completeness)}

\begin{proof}
Let $v_{\sigma(1)}, \dots, v_{\sigma(n)}$ be a Hamiltonian path. Wire
$s \to p^*_3$, energizing $b^*$, then wire $p^*_1 \to p^*_2$ to form a charged
loop, and disconnect $s$. Now $b^*$ is self-sustaining and $p^*_3$ is free and
energized.

Wire $p^*_3 \to p^L_{\sigma(1)}$, energizing $b_{\sigma(1)}$. For
$k = 1, \dots, n{-}1$, the edge $\{v_{\sigma(k)}, v_{\sigma(k+1)}\} \in E$
permits the wire $p^R_{\sigma(k)} \to p^L_{\sigma(k+1)}$, which energizes the
next board. No source is connected and every board is energized, so
$\mathcal{I}$ is winnable.
\end{proof}

\subsection*{Proof of Lemma~\ref{lem:sound} (Soundness)}

\begin{proof}
Consider a winning final configuration $W$: no source is connected, yet every
board is energized. With no sources, all energy originates from charged loops.
Recall (proof of Lemma~\ref{lem:three}) that any charged loop must contain a
board with at least three sockets. The only such board is $b^*$, so every
charged loop uses an anchor socket. Since $p^*_3$ is not connectable to
$p^*_1$ or $p^*_2$, the only loop on $b^*$ is $\{p^*_1, p^*_2\}$. Hence it is
the sole source of energy, and the node boards carry no loop of their own.

Every node board is therefore energized by a directed path of wires and
implicit edges starting at $b^*$. The only wire leaving $b^*$ goes from
$p^*_3$ (its single free socket) to one node board. Each node board has
exactly two sockets, each holding at most one wire, so once energy enters a
board through one socket it can leave only through the other. The energized
region is thus a single simple path
\[
  p^*_3 \to b_{\sigma(1)} \to b_{\sigma(2)} \to \cdots
\]
visiting distinct node boards. As all $n$ boards are energized, the path
visits every one, so $\sigma$ is a permutation of $\{1, \dots, n\}$. Each step
$b_{\sigma(k)} \to b_{\sigma(k+1)}$ uses a wire, which exists only when
$\{v_{\sigma(k)}, v_{\sigma(k+1)}\} \in E$. Hence
$v_{\sigma(1)}, \dots, v_{\sigma(n)}$ is a Hamiltonian path in $G$.
\end{proof}

\subsection*{Proof of Lemma~\ref{lem:anchored}}

\begin{proof}
Membership is immediate (a path is a certificate). For hardness we reduce from
the following form of grid-graph Hamiltonian path, which is
\NP-complete~\cite{itai}: given a grid graph $G$ and two vertices $s, t$ on
the outer boundary of its embedding, decide whether $G$ has a Hamiltonian
$s$--$t$ path. Boundary terminals are what the standard construction produces. The only feature of the instance we use is that $s$ and $t$ each have an empty
orthogonally adjacent exterior cell. Let $G$ sit inside a bounding box. From
$s$ route a self-avoiding path of new grid vertices: step first into that
empty neighbour, then move through exterior cells that stay at least two cells
away from every vertex of $G$, out of the bounding box and up, ending in a tip
$r$ strictly above the box. Staying two cells clear makes the new vertices
adjacent only to one another (and to $s$ at the first step), so the arm is a
chain of degree-$\le 2$ vertices with a degree-one tip $r$ and creates no
other adjacency with $G$. Routing it above the box makes $r$ the strictly
highest vertex. Attach a like arm to $t$ ending in a tip $r'$, routed through
the exterior on the opposite side and kept two cells clear of the first arm so
the two do not interfere. The augmented graph $G'$ is a grid graph. Because
each arm is a chain of degree-$\le 2$ vertices, any Hamiltonian path of $G'$
must traverse each arm end to end, and since $r, r'$ have degree one they are
forced to be the two endpoints. Deleting the arms turns a Hamiltonian path of
$G'$ into a Hamiltonian $s$--$t$ path of $G$, and conversely any such path
extends through the arms. Hence $G'$, with $r$ highest and of degree one, has
a Hamiltonian path with endpoint $r$ iff $G$ has one from $s$ to $t$.
\end{proof}

\subsection*{Proof of Theorem~\ref{thm:fixeddim}}

\begin{figure}[h]
  \centering
  \begin{subfigure}[t]{0.45\textwidth}
    \centering
    \includegraphics[width=\linewidth]{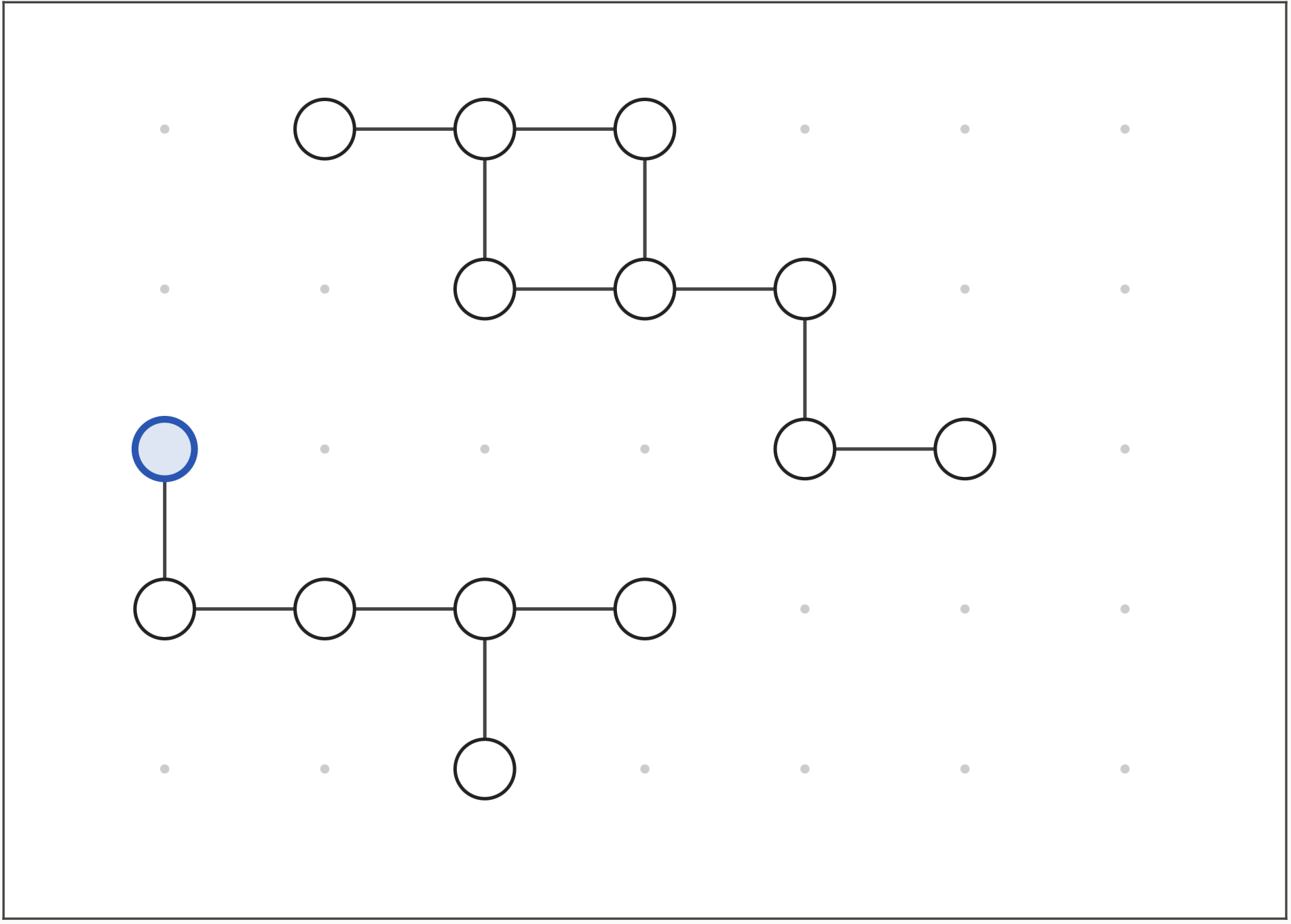}
    \caption{Arbitrary grid graph on which a Hamiltonian Path must begin at
    the \textcolor{blue}{highlighted} node. An edge exists between adjacent
    nodes on the grid.}
  \end{subfigure}\hfill
  \begin{subfigure}[t]{0.45\textwidth}
    \centering
    \includegraphics[width=\linewidth]{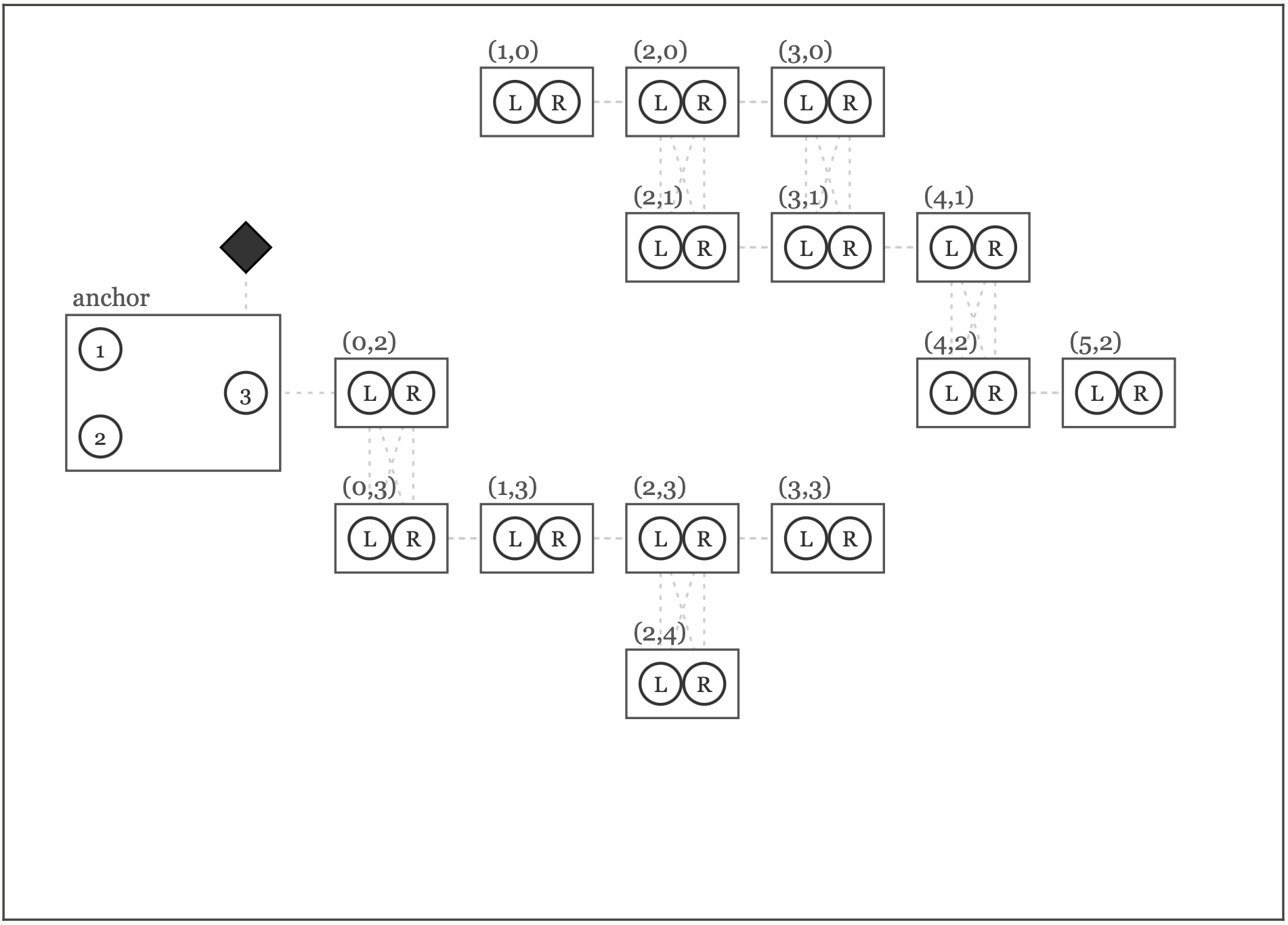}
    \caption{Corresponding game instance, where the dotted lines denote
    connectability governed by a distance threshold.}
  \end{subfigure}
  \caption{The $\R^2$ reduction of Theorem~\ref{thm:fixeddim}, drawn
  schematically. (a) A grid graph whose Hamiltonian path must end at the
  highlighted degree-one terminal $r$. (b) The game instance: each vertex
  becomes a two-socket board at its lattice point, so two boards are
  connectable (dotted lines, distance $\le d$) iff the vertices are
  grid-adjacent. The anchor board and source sit in an empty region beside $r$
  and reach only $r$'s board. In the reduction proper
  (Theorem~\ref{thm:fixeddim}) $r$ is placed as the highest vertex with the
  anchor in the empty half-plane above it. Further examples are at
  \url{https://infinite-energy-peer-review.netlify.app}.}
  \label{fig:r2example}
\end{figure}

\begin{proof}[sketch]
It suffices to treat $k = 2$, since any $\R^2$ realization embeds in $\R^k$ by
padding with zeros. We reduce from \textsc{Anchored Hamiltonian Path} on a
grid graph $(G, r)$ with $\deg(r) = 1$ (Lemma~\ref{lem:anchored}). Fix
$\varepsilon = 0.01$ and threshold $d = 1.05$.

\emph{Node boards.} For each grid vertex $v = (a, b)$ create a board
$B_v = \{(a - \varepsilon, b), (a + \varepsilon, b)\}$. For adjacent vertices,
horizontal or vertical, \emph{every} socket pair is within $d$: the farthest
such pair is at distance $1 + 2\varepsilon = 1.02 \le d$. For diagonal
vertices the nearest pair is at distance
$\sqrt{(1 - 2\varepsilon)^2 + 1} \approx 1.40 > d$, and farther pairs are
larger. Thus two node boards are connectable iff their vertices are
grid-adjacent, and, as in Section~\ref{sec:nphard}, all four socket pairings
between adjacent boards are available, so a Hamiltonian path can use distinct
incoming and outgoing sockets at every internal board. (Intra-board pairings
are immaterial: by Lemma~\ref{lem:three} no two-socket board can sustain a
charged loop.) Every node board has diameter $2\varepsilon$.

\emph{Anchor and source.} By Lemma~\ref{lem:anchored} embed $G$ with $r$ its
strictly highest vertex and place $r$ at the origin, so every other vertex has
$y \le -1$ and the region $y \ge 1$ is empty. Put $p^*_3 = (0, 1)$,\quad
$\sigma = (0.9, 1)$,\quad $p^*_1 = (-2, 1)$,\quad $p^*_2 = (-2, 2)$, with
anchor board $b^* = \{p^*_1, p^*_2, p^*_3\}$ and source $\sigma$. One checks:
$\|p^*_1 - p^*_2\| = 1 \le d$ (the anchor's loop).
$\|p^*_3 - (\pm\varepsilon, 0)\| \approx 1.00005 \le d$, so $p^*_3$ connects
to $B_r$, while $\|p^*_3 - p^*_1\| = 2$ and $\|p^*_3 - p^*_2\| = \sqrt{5}$ are
both $> d$ and $\|\sigma - (\pm\varepsilon, 0)\| \approx 1.34 > d$, so $p^*_3$
is not wireable to $p^*_1, p^*_2$ and $\sigma$ reaches only $p^*_3$. Every
other node board lies at $y \le -1$, at distance $> d$ from all four points.

This matrix is realized in $\R^2$ and matches the construction of
Section~\ref{sec:nphard} with the single change that $p^*_3$ is wireable only
to $B_r$. Hence the only anchor loop is $\{p^*_1, p^*_2\}$, energy leaves $b^*$
only through $p^*_3$ into $B_r$, and the degree-two node boards force a single
simple path covering them all. The completeness and soundness arguments of
Theorem~\ref{thm:nphard} then go through, now with the path forced to start at
$r$: $\mathcal{I}$ is winnable iff $G$ has a Hamiltonian path with endpoint
$r$. See Fig.~\ref{fig:r2example} for an example reduction.
\end{proof}

\subsection*{Proof of Theorem~\ref{thm:tc0}}

\begin{proof}
Let $x \in \{0,1\}^m$ with $s = \sum_{i=1}^m x_i$ ones. Map $x$ to the
instance with one board per bit:
\[
  b_i =
  \begin{cases}
    \{p_{i,1}, p_{i,2}, p_{i,3}\}, & x_i = 1,\\
    \{p_{i,1}\}, & x_i = 0,
  \end{cases}
  \qquad
  P = \bigcup_{i=1}^m b_i, \qquad B = \{b_1, \dots, b_m\}.
\]
Here $|B| = m$ and $|P| = 3s + (m - s) = m + 2s$, giving
\[
  |P| \ge 2|B| \iff m + 2s \ge 2m \iff s \ge \tfrac{m}{2}.
\]
Moreover, when $s \ge m/2$ and $m > 0$ some bit is $1$, so some board has
three sockets and the second condition of Theorem~\ref{thm:char} holds
automatically. When $s < m/2$ the size condition already fails. Thus the
instance is accepted iff $x$ has a majority of ones. Each output board depends
on a single input bit, so the map is a projection, and every projection is an
\AC{} reduction. Since \MAJ{} is \TC-complete under \AC{} Turing
reductions~\cite{vollmer}, and \UIE{} $\in \mTC$
(Proposition~\ref{prop:tc0}), the problem is \TC-complete under \AC{} Turing
reductions.
\end{proof}

\section{Proof of Theorem~\ref{thm:realize}}\label{app:realize}

\begin{proof}
Write $A = M$ and pick $\lambda > \max\{1, -\lambda_{\min}(A)\}$, so that
$B := \lambda I + A$ is positive semidefinite. Then $B$ is a Gram matrix: there
are vectors $z_1, \dots, z_n$ with $z_i^\top z_j = B_{ij}$. Hence
$\|z_i\|^2 = \lambda$ for all $i$, and $z_i^\top z_j = A_{ij} \in \{0,1\}$ for
$i \neq j$, so
\[
  \|z_i - z_j\|^2 = 2\lambda - 2A_{ij} =
  \begin{cases}
    2\lambda - 2, & A_{ij} = 1,\\
    2\lambda, & A_{ij} = 0.
  \end{cases}
\]
Choosing threshold $d^2 = 2\lambda - 1$ separates the two cases:
$A_{ij} = 1 \iff \|z_i - z_j\| \le d$.

It remains to reduce the dimension. Translate by anchoring $z_n$: set
$y_i = z_i - z_n$, so $y_n = 0$ and $\|y_i - y_j\| = \|z_i - z_j\|$ for all
$i, j$. The threshold graph is unchanged. The $y_i$ lie in
$\operatorname{span}\{y_1, \dots, y_{n-1}\}$, a space of dimension at most
$n - 1$. Identifying it with $\R^{n-1}$ via an orthonormal basis realizes $M$
in $\R^{n-1}$. For further background, see standard sources such
as~\cite{hlineny,linial}.
\end{proof}

\begin{remark}
Theorem~\ref{thm:realize} is an existence statement about the matrix, so
Corollary~\ref{cor:geom} classifies the variant whose input is a matrix
promised to be realizable. The coordinates it produces are real and need not
have polynomial bit-length. The fixed-dimension reduction of
Theorem~\ref{thm:fixeddim} is different: it outputs explicit rational
coordinates (lattice points offset by $\varepsilon = 0.01$, threshold
$d = 1.05$) of polynomial precision, so it is a reduction to the
explicit-coordinate problem.
\end{remark}

\section{Example of the \texorpdfstring{$\R^2$}{R\^{}2} reduction}\label{app:example}

\section{Extension: the color-constrained variant, and open directions}\label{app:color}

This appendix records the colored extension and the questions it raises. The
only complexity result here is an easy inherited hardness, and the more
interesting questions are open. We give the model, note \NP-hardness, and
explain why the colorless membership argument does not carry over.

\begin{figure}[h]
  \centering
  \includegraphics{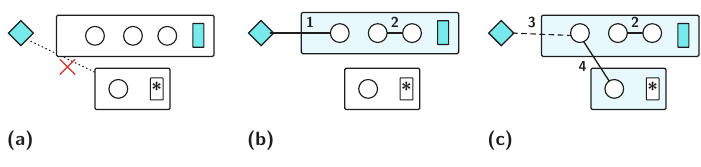}
  \caption{A colored play. (a) The initial configuration. The rectangle on the
  top right of each board shows its color requirement $F$, and the source
  cannot reach the lower one-socket board because $M$ forbids the wire (too
  far away). (b) In step 1 the source is wired into the board whose
  requirement is blue. In step 2 a loop is formed on that board. (c) In step 3
  (dotted) the source is disconnected and in step 4 the now self-sustaining
  board energizes the lower board through a cross-board wire. Every board
  meets its requirement, so the instance is won.}
  \label{fig:colorplay}
\end{figure}

We extend the colorless game of Section~\ref{sec:intro} with colors. The
mechanic is the additive mixing of light: combining colored beams only ever
adds primaries, never removes them, so mixing is modeled by union and named
colors are abbreviations for subsets of primaries.

\paragraph{Color algebra.} Fix a finite set of \emph{primary colors} $C'$. The
figures use the three additive primaries red, green, and blue. A \emph{color}
is a nonempty subset of $C'$, and $\Col = \mathcal{P}(C') \setminus
\{\emptyset\}$. Colors mix by union, $A \oplus B = A \cup B$, an operation
that is commutative, associative, and idempotent. The familiar names label
elements of $\Col$:
\begin{gather*}
  \{\mathsf{red}, \mathsf{blue}\} = \mathsf{purple}, \quad
  \{\mathsf{red}, \mathsf{green}\} = \mathsf{yellow},\\
  \{\mathsf{blue}, \mathsf{green}\} = \mathsf{cyan}, \quad
  C' = \mathsf{white}.
\end{gather*}

We adjoin a distinguished symbol $\star \notin \Col$ denoting the
\emph{any-color} requirement.

\paragraph{Instance.} Extending Definition~\ref{def:instance}, a \emph{colored
instance} $\mathcal{I} = (P, S, B, M, \chi, F)$ adds to a colorless instance
$(P, S, B, M)$ two maps:
\begin{itemize}
  \item $\chi : S \to \Col$ assigns a color to each power source.
  \item $F : B \to \Col \cup \{\star\}$ assigns a color requirement to each
  board.
\end{itemize}

\begin{figure}[t]
  \centering
  \includegraphics{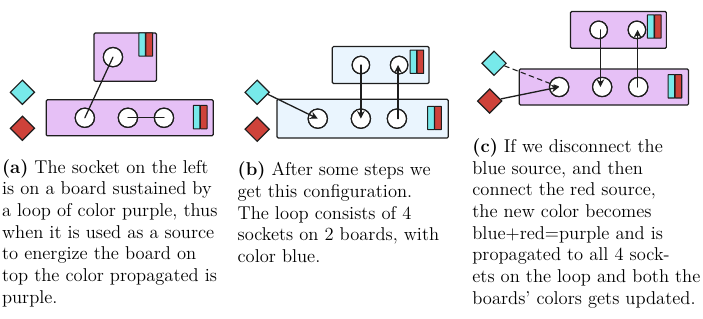}
  \caption{Color mixing and propagation on a loop. (a) A socket on a board
  sustained by a purple loop exports purple. (b) A blue loop spanning four
  sockets on two boards. (c) Disconnecting the blue source and then connecting
  a red source turns the loop purple ($\mathsf{blue} \oplus \mathsf{red}$),
  and the new color propagates to all four sockets and both boards. Panel (c)
  shows why the \emph{order} of source connections and disconnections matters
  in the colored game.}
  \label{fig:colormix}
\end{figure}
\noindent Configurations, wires, implicit edges, loops, and the energized sets $\en_t$
are exactly as in Section~\ref{sec:intro} and
Appendix~\ref{app:membership}. Colors are on top of energization and do not
affect which sockets are energized.

\paragraph{Color propagation.} Fix a play $E_0, \dots, E_T$ (energization is
as in Appendix~\ref{app:membership}). Color accumulates along the play by the
same flow, and transmission is \emph{gated by energization}: a socket passes
its color along an edge only while it is energized. For a socket $p \in P$,
its color $\col_t(p)$ is the least solution of
\[
  \col_t(p) \;=\; \bigoplus \Bigl\{\, \chi(s) \Bigm| s \in S,\ s \text{ reaches }
  p \text{ in } G'(E_{t'}),\ t' \le t \,\Bigr\}
  \;\oplus\;
  \bigoplus_{\substack{(u,p) \in E_{t'} \cup E_B,\ t' \le t\\ u \in \en_{t'}}}
  \col_{t'}(u),
\]
i.e.\ the union of the colors of every source whose energy has reached $p$ at
any point, propagated forward through wires, implicit edges, and around
charged loops. We take this \emph{permanent-memory} model: a socket keeps
every primary it has ever received, so $\col_t(p)$ never shrinks, but by the
gate it passes color on only while energized. For definiteness
$\col_t(s) = \chi(s)$ for a source $s$, and $\col_t(p) = \emptyset$ for a
socket never energized by step $t$. Retaining a color while dark thus differs
from transmitting one: on two singleton boards $\{a\}, \{b\}$, wiring a red
source to $a$, unplugging it, then adding $a \to b$ leaves $b$ colorless,
because $a$ is dark when $a \to b$ is present. A reset-on-de-energization
variant is a different model with a different recurrence. The \emph{effective
color} of an energized board $b$ at time $t$ is
\[
  \col'_t(b) \;=\; \bigoplus \bigl\{\, \col_t(p) \bigm| p \in b,\ p \text{
  energized in } E_t \,\bigr\}.
\]

See Fig.~\ref{fig:colorplay} and~\ref{fig:colormix} for examples.

\paragraph{Board satisfaction and the decision problem.} A board $b$ is
\emph{satisfied} at $t$ if it is energized in $E_t$ and its requirement is met
exactly:
\[
  F(b) = \star \qquad\text{or}\qquad \col'_t(b) = F(b).
\]

\paragraph{Colored Infinite Energy} Given a colored instance
$\mathcal{I} = (P, S, B, M, \chi, F)$, does there exist a play whose final
configuration $E_T$ connects no source yet leaves every board satisfied?

\paragraph{Hardness.}

\begin{proposition}\label{prop:colorhard}
Colored \IE{} is \NP-hard.
\end{proposition}

\begin{proof}
Colorless \IE{} is the special case in which color is vacuous. Given a
colorless instance $(P, S, B, M)$, take a single primary $C' = \{\bullet\}$,
set $\chi(s) = \{\bullet\}$ for every $s \in S$, and $F(b) = \star$ for every
$b \in B$. Then every energized board is satisfied, so the colored instance is
winnable iff the colorless one is. The map is identity on $(P, S, B, M)$ and
constant on $\chi, F$, hence a polynomial-time (indeed projection) reduction.
\NP-hardness follows from Theorem~\ref{thm:nphard}.
\end{proof}

\paragraph{Open problems.} The upper bound is open, and for a sharper reason
than in the colorless case. There, whether a board is energized does not
depend on the order in which sources are used, which is what the short-play
route (Theorem~\ref{thm:np}, itself conditional) exploits. With colors the
requirement $\col'(b) = F(b)$ is an \emph{exact} match while color only
accumulates (Fig.~\ref{fig:colormix}), so a board can be over-colored and
never corrected. Meeting every requirement at once then constrains the order
in which sources are connected and disconnected across the whole play, which
the final configuration does not record. We leave three questions. (i) Is
colored \IE{} in \NP? (ii) Is it hard for a class above \NP? It is in \PSPACE{}
by the state-space bound of Theorem~\ref{thm:pspace}, which applies verbatim
once colors are added to the state. (iii) Does the reset-on-de-energization
model differ in complexity from the permanent-memory one above?

\end{document}